\documentclass[sigconf, nonacm, pdfa]{acmart}
\usepackage{todonotes}
\usepackage{xspace}
\usepackage[a-2b]{pdfx}
\usepackage{balance}

\newcommand\vldbdoi{10.14778/3611540.3611569}
\newcommand\vldbpages{3848-3860}
\newcommand\vldbvolume{16}
\newcommand\vldbissue{12}
\newcommand\vldbyear{2023}
\newcommand\vldbauthors{\authors}
\newcommand\vldbtitle{\shorttitle} 
\newcommand\vldbavailabilityurl{URL_TO_YOUR_ARTIFACTS}
\newcommand\vldbpagestyle{empty} 

\def\code#1{\texttt{{\footnotesize#1}}}

\makeatletter
\renewcommand\paragraph{\@startsection{paragraph}{4}{\z@}%
                                     {-3.25ex\@plus -1ex \@minus -.2ex}%
                                     {1.5ex \@plus .2ex}%
                                     {\normalfont\normalsize\bfseries}}
\makeatother

\begin{document}
\title{PyTorch FSDP: Experiences on Scaling Fully Sharded Data Parallel}

% \author{Yanli Zhao \ \ \  Andrew Gu \ \ \  Rohan Varma \ \ \  Liang Luo \ \ \  Chien-Chin Huang \ \ \  Min Xu \\ 
% Less Wright \ \ \  Hamid Shojanazeri \ \ \   Myle Ott \ \ \   Sam Shleifer \ \ \  Alban Desmaison \ \ \  Can Balioglu \\ Pritam Damania \ \ \
% Bernard Nguyen \ \ \   Geeta Chauhan \ \ \   Yuchen Hao \ \ \   Shen Li\\}
% \affiliation{%
%   \institution{Meta AI}
% }
% \email{{yanlizhao, andgu, rvarm1, liangluo, chienchin, m1n}@meta.com}
% \email{{less, hamidnazeri}@meta.com, {myleott, sshleifer}@gmail.com,{albandes, balioglu}@meta.com}
% \email{pritam.damania@gmail.com, {bernardn, gchauhan, haoyc, shenli}@meta.com}

\author{Yanli Zhao}
\affiliation{%
  \institution{Meta AI}
}
\email{yanlizhao@meta.com}

\author{Andrew Gu}
\affiliation{%
  \institution{Meta AI}
}
\email{andgu@meta.com}

\author{Rohan Varma}
\affiliation{%
  \institution{Meta AI}
}
\email{rvarm1@meta.com}

\author{Liang Luo}
\affiliation{%
  \institution{Meta AI}
}
\email{liangluo@meta.com}

\author{Chien-Chin Huang}
\affiliation{%
  \institution{Meta AI}
}
\email{chienchin@meta.com}

\author{Min Xu}
\affiliation{%
  \institution{Meta AI}
}
\email{m1n@meta.com}

\author{Less Wright}
\affiliation{%
  \institution{Meta AI}
}
\email{less@meta.com}

\author{Hamid Shojanazeri}
\affiliation{%
  \institution{Meta AI}
}
\email{hamidnazeri@meta.com}

\author{Myle Ott}
\affiliation{%
  \institution{Meta AI}
}
\email{myleott@gmail.com}

\author{Sam Shleifer}
\affiliation{%
  \institution{Meta AI}
}
\email{sshleifer@gmail.com}

\author{Alban Desmaison}
\affiliation{%
  \institution{Meta AI}
}
\email{albandes@meta.com}

\author{Can Balioglu}
\affiliation{%
  \institution{Meta AI}
}
\email{balioglu@meta.com}

\author{Pritam Damania}
\affiliation{%
  \institution{Meta AI}
}
\email{pritam.damania@gmail.com}

\author{Bernard Nguyen}
\affiliation{%
  \institution{Meta AI}
}
\email{bernardn@meta.com}

\author{Geeta Chauhan}
\affiliation{%
  \institution{Meta AI}
}
\email{gchauhan@meta.com}

\author{Yuchen Hao}
\affiliation{%
  \institution{Meta AI}
}
\email{haoyc@meta.com}

\author{Ajit Mathews}
\affiliation{%
  \institution{Meta AI}
}
\email{amath@meta.com}

\author{Shen Li}
\affiliation{%
  \institution{Meta AI}
}
\email{shenli@meta.com}

%
% The abstract is a short summary of the work to be presented in the
% article.
\begin{abstract}
It is widely acknowledged that large models have the potential to deliver superior performance across a broad range of domains. Despite the remarkable progress made in the field of machine learning systems research, which has enabled the development and exploration of large models, such abilities remain confined to a small group of advanced users and industry leaders, resulting in an implicit technical barrier for the wider community to access and leverage these technologies. In this paper, we introduce PyTorch Fully Sharded Data Parallel (FSDP) as an industry-grade solution for large model training. FSDP has been closely co-designed with several key PyTorch core components including Tensor implementation, dispatcher system, and CUDA memory caching allocator, to provide non-intrusive user experiences and high training efficiency. Additionally, FSDP natively incorporates a range of techniques and settings to optimize resource utilization across a variety of hardware configurations. The experimental results demonstrate that FSDP is capable of achieving comparable performance to Distributed Data Parallel while providing support for significantly larger models with near-linear scalability in terms of TFLOPS.
\end{abstract}

\maketitle

\pagestyle{\vldbpagestyle}
\begingroup\small\noindent\raggedright\textbf{PVLDB Reference Format:}\\
\vldbauthors. \vldbtitle. PVLDB, \vldbvolume(\vldbissue): \vldbpages, \vldbyear.\\
\href{https://doi.org/\vldbdoi}{doi:\vldbdoi}
\endgroup
\begingroup
\renewcommand\thefootnote{}\footnote{\noindent
This work is licensed under the Creative Commons BY-NC-ND 4.0 International License. Visit \url{https://creativecommons.org/licenses/by-nc-nd/4.0/} to view a copy of this license. For any use beyond those covered by this license, obtain permission by emailing \href{mailto:info@vldb.org}{info@vldb.org}. Copyright is held by the owner/author(s). Publication rights licensed to the VLDB Endowment. \\
\raggedright Proceedings of the VLDB Endowment, Vol. \vldbvolume, No. \vldbissue\ %
ISSN 2150-8097. \\
\href{https://doi.org/\vldbdoi}{doi:\vldbdoi} \\
}\addtocounter{footnote}{-1}\endgroup

\ifdefempty{\vldbavailabilityurl}{}{
\vspace{.3cm}
\begingroup\small\noindent\raggedright\textbf{PVLDB Artifact Availability:}\\
The source code, data, and/or other artifacts have been made available at \url{https://github.com/pytorch/pytorch/blob/main/torch/distributed/fsdp/fully_sharded_data_parallel.py/}.
\endgroup
}

\section{Introduction}
% To-do: Establish that FSDP is acronym for Fully Sharded Data Parallel
% We generally refer to FSDP to describe the concept as a whole. We specify FullyShardedDataParallel if referring to the Python class itself.

The magnitude of neural network models is growing at an unprecedented rate, facilitating breakthroughs across a wide spectrum of domains. Upon inception, the 175-billion-parameter GPT-3~\cite{gpt3} model set a new record for almost all Natural Language Processing tasks. The product applications constructed on top of GPT models~\cite{chatgpt} have quickly demonstrated their potential to revolutionize the entire industry. Modern large scale recommendation models~\cite{mudigere2021high, DHEN} can reach beyond 1 trillion parameters, replete with rapidly growing dense layer components. These models power applications that serve multi-billions of users every day. As large neural networks continue to push the limits of science and technology, an industry-grade tool to simplify the training of such models with high efficiency would help expedite the progress.

In recent years, the community has introduced and investigated numerous advanced methodologies to enlarge neural network models. Pipeline parallelism~\cite{gpipe, kim2020torchgpipe, pipetransformer, li2021terapipe, narayanan2019pipedream} partitions a model instance into stages and distributes stages across multiple devices, where activations and gradients are communicated across stage boundaries. Tensor parallelism~\cite{flexflow, narayanan2021efficient, gspmd, yuan2021oneflow} shards model parameters, conducts partial computation on individual devices and communicates activations at required layer boundaries. Zero-Redundancy parallelism~\cite{zero, zero:offload, xu2020automatic} shards parameters as well but communicates parameters on-demand to recover their unsharded form and executes the model as if it were replicated on every device. The aforementioned techniques have served as the fundamental building blocks to enable the training of large neural networks across various applications. Nevertheless, two challenges still persist. Firstly, some of these methods are tightly integrated with specific model architectures, which hinder them from being utilized as a generic solution for training large models. Secondly, some of these techniques are built on top of rapidly-evolving internal interfaces of underlying machine learning frameworks, which become vulnerable to changes in framework implementations. Therefore, it is more robust and efficient to have a native solution co-designed with the core functionalities of machine learning frameworks. Additionally, constructing such a solution in a composable and customizable manner could potentially facilitate the community's future innovations as well.

This paper presents PyTorch~\cite{pytorch:nips:2019} Fully Sharded Data Parallel (FSDP), which enables the training of large-scale models by sharding model parameters. The FSDP algorithm is motivated by the ZeroRedundancyOptimizer~\cite{zero, zero:offload} technique from DeepSpeed but with a revised design and implementation that is aligned with the other components of PyTorch. FSDP breaks down a model instance into smaller units and then flattens and shards all of the parameters within each unit. The sharded parameters are communicated and recovered on-demand before computations, and then they are immediately discarded afterwards. This approach ensures that FSDP only needs to materialize parameters from one unit at a time, which significantly reduces peak memory consumption. The design and implementation of FSDP faces the following challenges.

\begin{itemize}
    \item \textbf{User Experience} is critical for achieving broad adoption. When working on prior PyTorch distributed training features such as \code{DistributeDataParallel} (DDP)~\cite{ddp}, we observed that aligning the user experience of distributed training with that of local training can significantly lower the learning barrier. Techniques like DDP require the model to be replicated on every device, which implies that the entire model can be constructed on the target device. However, although FSDP can easily adopt DDP's API design, large models might not fit into one GPU device and therefore cannot even be initialized efficiently. 
    \item \textbf{Hardware Heterogeneity} often exists in modern GPU clusters, whereby interconnects are partitioned into high-bandwidth islands within each machine and low-bandwidth mesh across machines. Additionally, there may be further hierarchical structures at the rack or pod levels. Consequently, the design of FSDP must accommodate such heterogeneity and optimize accordingly. 
    \item \textbf{Resource Utilization} is usually tightly linked with capital and operational expenditures, especially for companies that depend on large GPU clusters to power their mission-critical systems. To ensure that GPU devices remain fully utilized during distributed training, it is essential to minimize downtime caused by non-computational operations.
    \item \textbf{Memory Planning} plays a crucial role in large model training. PyTorch makes GPU memory block allocation efficient and transparent through caching. Frequent memory defragmentations can significantly slow down training, which becomes particularly acute when working with large models. In such scenarios, practitioners typically seek to saturate GPU memory as much as possible to accommodate the largest batches or models. However, operating near GPU memory capacity significantly increases the chance to trigger defragmentations.
\end{itemize}

FSDP tackles the aforementioned challenges through a variety of techniques. Firstly, to improve user experience, FSDP introduces deferred initialization that allows users to create a model instance on a dummy device and record operations invoked during initialization. Then, the model can be initialized and sharded unit by unit by replaying the recorded operations on a real GPU device. With this technique, FSDP can provide similar user experiences as local training, while effectively scaling large models. Secondly, FSDP offers configurable sharding strategies that can be customized to match the physical interconnect topology of the cluster to handle hardware heterogeneity. Thirdly, although parameter sharding design inevitably inserts communications, which might block computations and introduces bubbles during execution, FSDP can squeeze out bubbles using an abundant set of tools to aggressively overlap communication with computation through operation reordering and parameter prefetching. Lastly, FSDP optimizes memory usage by prudently restricting the amount of blocks allocated for inflight unsharded parameters and suspending CPU execution if necessary.

We evaluated the performance of FSDP on various models including popular language models and recommendation system models, utilizing up to 512 80GB A100 GPUs. The experiments showed that FSDP can achieve similar performance to that of DDP on small models. Beyond that FDSP can facilitate significantly larger models with near-linear scalability in terms of TFLOPS. FSDP is currently a beta feature as of PyTorch 2.0 release, and has been battle-tested by both industrial and research applications. 

To simplify presentation, the rest of this paper uses FSDP to refer to the techniques in general and \code{FullyShardedDataParallel} to denote the Python implementation. The remainder of the paper is organized as follows. Section~\ref{sec:background} introduces background on some popular distributed training techniques. Section~\ref{sec:design} and Section~\ref{sec:implementation} elaborate system design and implementation details. Evaluations are presented in Section~\ref{sec:evaluation}. Section~\ref{sec:related} surveys related work, and Section~\ref{sec:discussion} discusses topics related to FSDP but falls outside of FSDP core. Finally, Section~\ref{sec:conclusion} concludes the paper.

\section{Background}\label{sec:background}

PyTorch~\cite{pytorch:nips:2019} has emerged as a fundamental cornerstone for a plethora of machine learning endeavors. PyTorch stores values in \code{Tensor} objects, which are versatile n-dimensional arrays featuring a rich set of data manipulation operations. Every \code{Tensor} object has an associated storage that is allocated on a specific device. When \code{Tensor}s only represent simple transformations such as \code{reshape} and \code{split}, they can share the same underlying storage.  Each \code{Module} describes a transformation from input to output values, and its behavior during the forward pass is specified by its \code{forward} member function. Such a module may feature \code{Tensor} objects as parameters, with the \code{Linear} module being an example that contains both \code{weight} and \code{bias} parameters. During the forward pass, the \code{Linear} module applies these parameters to the input to produce the output by means of multiplication and addition operations, respectively.

As both the data size and model complexity continue to escalate at a staggering pace, the need for an industry-grade distributed training framework becomes increasingly imperative for applications built on top of PyTorch. This section elucidates the trajectory of PyTorch's distributed training capabilities.

\subsection{Model Replication}

Model replication approaches are designed to tackle high-volume datasets by scaling out and distributing computations across multiple devices. \code{DistributedDataParallel} (DDP)~\cite{ddp} is the first end-to-end distributed training feature in PyTorch that falls into this category. DDP's adoption has been extensive, spanning both the academic and industrial domains. 

DDP maintains a model replica on each device and synchronizes gradients through collective \code{AllReduce} operations in the backward pass, thereby ensuring model consistency across replicas during training. To expedite training, DDP overlaps gradient communication with backward computation, facilitating concurrent workload executions on diverse resources. However, one conspicuous limitation is that DDP requires all model parameters, gradients, and optimizer states to fit in the memory of one GPU device. Consequently, DDP is inadequate for supporting large models, which are critical for cutting-edge machine learning breakthroughs. For example, when training models with more than one billion parameters using a 40GB GPU device, DDP will likely encounter out-of-memory errors on each device.

\subsection{Model Partitioning}

As the size of models grow, they may no longer fit in a single GPU device. In such cases, a viable solution is to partition the model into smaller components and distribute them across multiple devices. Both pipeline parallelism~\cite{gpipe} and Tensor RPC~\cite{rpc} are along this direction. Pipeline parallelism involves breaking a sequence of layers into stages and feeding inputs to different stages in a pipelined fashion to optimize resource utilization. On the other hand, Tensor RPC provides a lower-level toolkit that enables arbitrary computations to be executed on remote devices. While both techniques are capable of scaling large models across multiple devices, they either limit the model to a sequence of stages or require modifications to the model authoring code to insert remote computations, which can pose a significant obstacle to users' adoption. Moreover, many industrial training infrastructures only support the single-program multi-data paradigm, which necessitates a simpler entry point to handle large models.

\subsection{Model Sharding}

In addition to partitioning, sharding the parameters of a model can also help reduce its memory footprint and support models with sizes beyond the memory capacity of a single GPU device. After sharding models, each rank only holds a shard of the model parameters, which prevents it from performing the same computations as local training. To guarantee correctness, the training process needs to employ one or both of the following techniques:

\begin{itemize}
    \item Perform computations with parameter shards and \textbf{communicate activations} accordingly. With this approach, ranks never need to fully materialize any parameter. However, each communication will appear in the critical path as it is inserted between two consecutive and dependent computation operations. As a result, this communication cannot easily overlap with computations, unless non-dependent computations or computations from other iterations can be re-ordered to overlap with communication.
    \item Perform the same computation as local training by \textbf{communicating parameter} on-demand before computations. Since parameter communications do not have any data dependency on preceding computations, they can overlap with the preceding computations performed in the same forward or backward pass. However, this approach requires that the on-demand communicated parameters could be fully materialized and could fit in the memory of a single GPU device.
\end{itemize}

FSDP falls into the second category of communicating parameters. Based on our observations and experiments, this approach is sufficient to support the vast majority of large model applications today and in the near future. It is worth noting that if the requirement of fully materializing each parameter unit on GPU becomes a blocker, we can further combine both techniques to support such use cases.

\section{System Design}\label{sec:design}

Fully Sharded Data Parallel (FSDP) is capable of scaling to accommodate large models that may not fit in a single GPU device by sharding the dense parameters. More specifically, FSDP decomposes the model instance into smaller units and handles each unit independently. During forward and backward computation, FSDP only materializes unsharded parameters and gradients of one unit at a time, and otherwise, it keeps parameters and gradients sharded. Throughout the training loop, the optimizer states are kept sharded. The memory requirements for FSDP are proportional to the size of the sharded model plus the size of the largest fully-materialized FSDP unit.

\begin{figure}
    \centering
    \includegraphics[width=\linewidth]{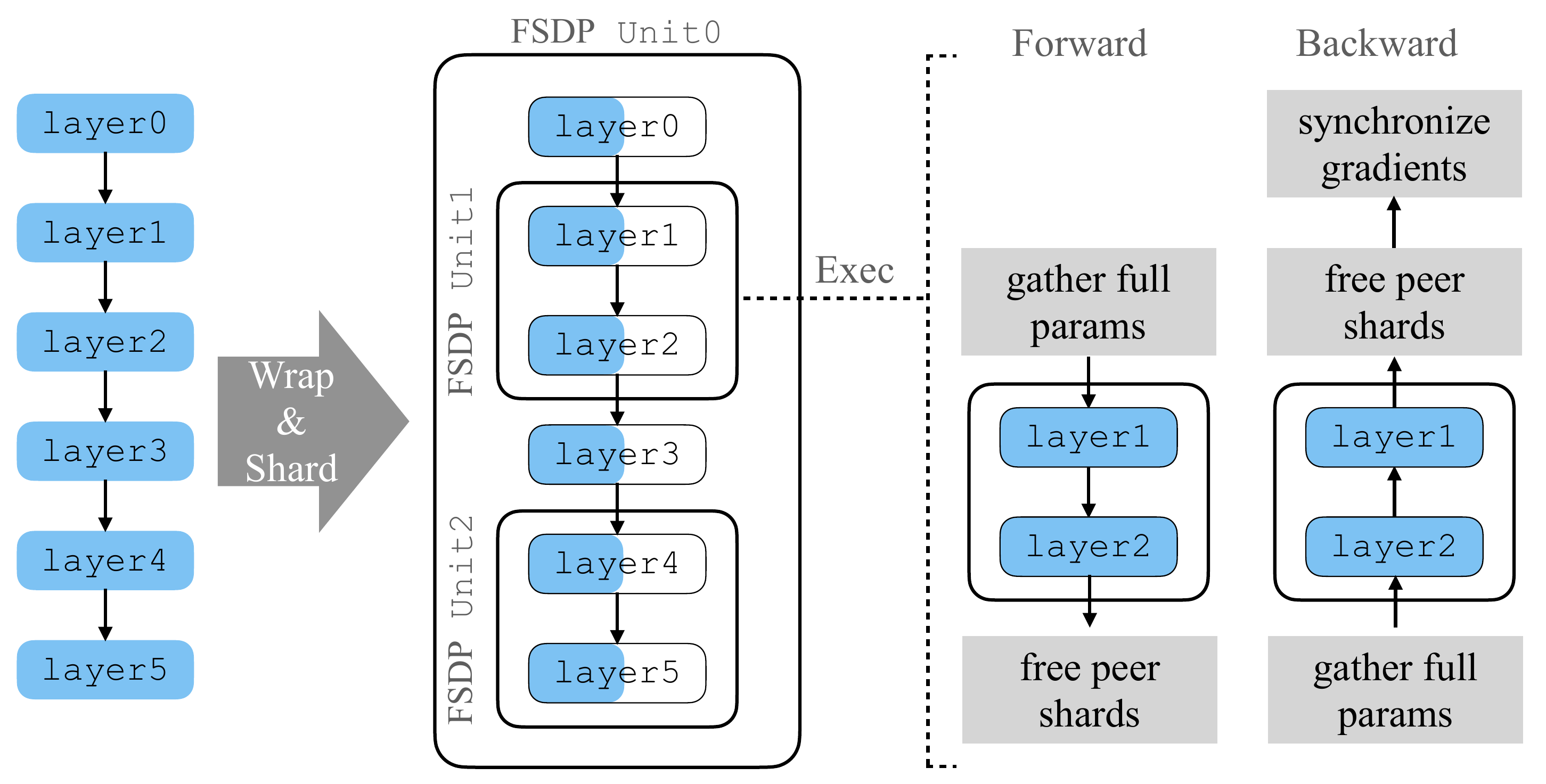}
    \caption{FSDP Algorithm Overview}
    \label{fig:overview}
\end{figure}

Figure~\ref{fig:overview} demonstrates the overall workflow using a simple six layer model. Suppose FSDP decomposes the model into three parts, namely, \code{[layer0, layer3]}, \code{[layer1, layer2]}, and \code{[layer4, layer5]}. The decomposition behavior can be controlled by user-defined functions. FSDP then wraps each of these three parts into one FSDP unit and shards parameters accordingly. To ensure correctness, FSDP needs to recover the unsharded parameters before corresponding computations. Let us consider FSDP \code{unit1} that contains \code{[layer1, layer2]} to explain this process. Before forward computation enters \code{layer1}, FSDP collects the unsharded parameters for \code{layer1} and \code{layer2} by gathering shards from other peer ranks. With the unsharded parameters, FSDP runs the local computation of those layers and then frees the peer shards it just collected to reduce memory footprint. Therefore, during the entire forward pass, FSDP only needs to fully materialize one unit at a time, while all other units can stay sharded. Similarly, during the backward computation, FSDP \code{unit1} recovers the unsharded parameters for \code{layer1} and \code{layer2} before backward reaches \code{layer2}. When the autograd engine finishes the backward computation of these two layers, FSDP frees the peer shards and launches \code{ReduceScatter} to reduce and shard gradients. Hence, after backward computation, each rank only keeps a shard of both parameters and gradients. 

FSDP offers a wide spectrum of optimizations and knobs to account for diverse model structures and hardware capabilities. The remainder of this section delves further into the intricacies of model initialization, sharding strategies, communication optimizations, and memory management, which are all critical components of FSDP's underlying design.

\subsection{Model Initialization}

Before the advent of FSDP, PyTorch mandated the full materialization of the entire model instance on one device. Although users can allocate different sub-modules to different devices, this would require modifying the model source code, which may not be feasible, particularly if model authors and application developers belong to different parties. To facilitate a smooth transition from local to distributed training, FSDP must effectively aid in the materialization and initialization of a massive model, which poses two challenges:

\begin{itemize}
    \item How to create a model instance without materializing any tensor storage, postponing initialization until a storage on a concrete device is attached to the tensor.
    \item How to ensure accurate initialization of model parameters in line with the user's implementation, even when the model is too large to fit on a single GPU. 
\end{itemize}

To overcome the first challenge, we have introduced a mechanism called \emph{deferred initialization}, which involves the allocation of model parameter tensors on a simulated or "fake" device. During this process, all initialization operations performed on the tensor are recorded. Subsequently, when the tensor is moved from the "fake" device to a GPU device, all recorded operations are automatically replayed. By adopting this technique, users can generate a model instance from any third-party library without allocating any GPU memory blocks, while still accurately capturing their parameter initialization implementations.

As illustrated in Figure~\ref{fig:overview}, once the FSDP has wrapped the model, it is evenly distributed across all GPUs, with each device holding only one shard in its memory. Therefore, in order to address the second challenge, each rank should ideally only materialize and initialize the shard that it owns. However, this is not always practical, since we cannot predict what initialization logic the user will implement in the model \code{init} method. The initialization logic may rely on having a unsharded parameter on the device, which makes it impossible to shard the initialization. Consequently, FSDP must prepare the unsharded parameters before executing \code{Tensor} initialization operations and simultaneously reduce the memory footprint. Given that sharding initialization is unsafe, FSDP applies the same approach as how it handles model forward and backward passes, \emph{i.e.}, initialize one FSDP unit at a time and shard the unit before moving on to the next one. When combined with deferred initialization, FSDP traverses the fake device model instance to decompose it into FSDP units, moves one unit to a GPU device at a time, and replays the recorded initialization operations for tensors in that FSDP unit.

%%%%%%%%%%%%%

\subsection{Sharding Strategies}\label{sec:sharding_strategy}

The \emph{sharding strategy} is an important element in FSDP that plays a significant role in determining the memory footprint and communication overhead.  FSDP offers a variety of sharding strategies, ranging from fully replicated to fully sharded. To generalize these sharding strategies, we introduce the \textit{sharding factor} $F$ as the number of ranks over which parameters are sharded. By setting the sharding factor to $1$, FSDP fully replicates the model and simplifies to vanilla data parallelism that uses \code{AllReduce} for gradient reduction. By setting the sharding factor equal to the number of devices (\emph{i.e.}, global world size $W$), FSDP fully shards the model, with each device only holding $\frac{1}{W}$ of the model. Hybrid sharding occurs when the sharding factor ranges between $1$ and $W$. The remainder of this section focuses on full sharding and hybrid sharding since the full replication strategy is similar to the existing DDP~\cite{ddp}.

\subsubsection{Full Sharding}\label{sec:full_shard}\hfill \break

The \emph{full sharding} strategy leads to the lowest memory footprint but incurs the most communication overhead, for example, \emph{full sharding} has 1.5x communication overhead and volume over DDP if using bandwidth optimal ring algorithm. Therefore, FSDP must carefully organize communications to maximize its efficiency under this strategy.

\begin{figure}

\begin{minipage}[c]{0.22\textwidth}
  \centering
  \includegraphics[width=\linewidth]{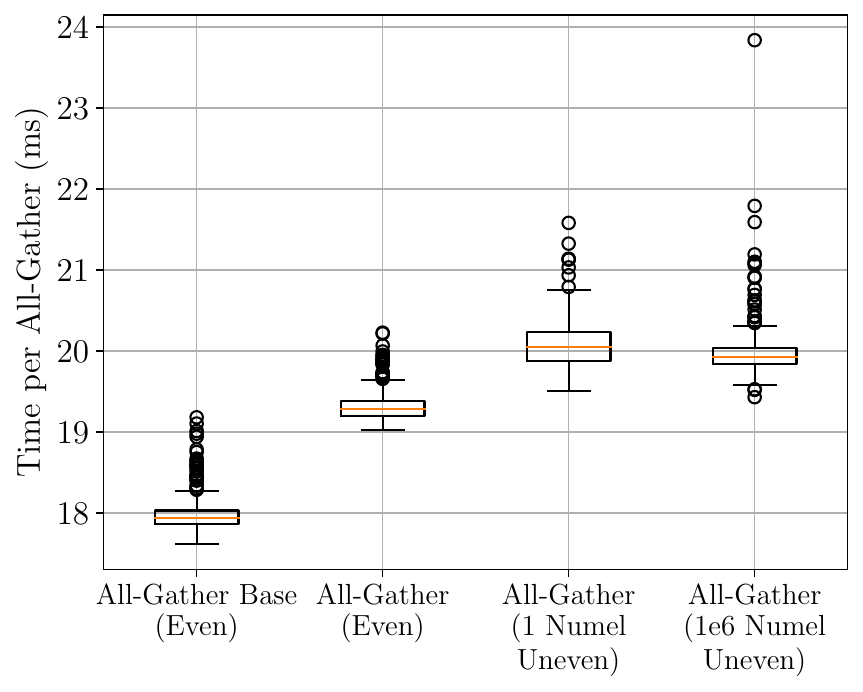}
  (a) Uneven Input Sizes
\end{minipage}
\begin{minipage}[c]{0.235\textwidth}
  \centering
  \includegraphics[width=\linewidth]{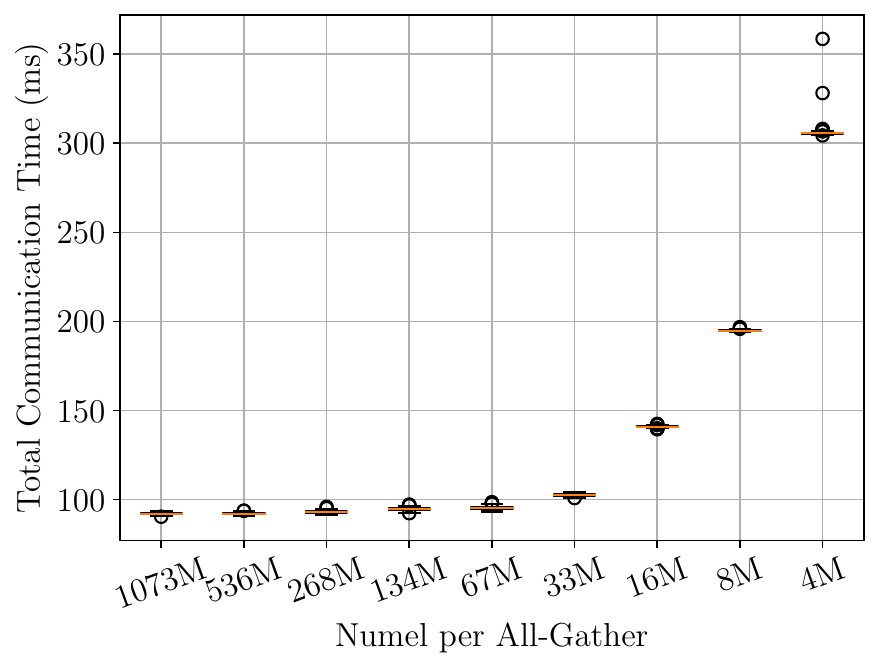}
  (b) Reducing Input Size
\end{minipage}
\caption{Communication Efficiency vs. Input Size}\label{fig:comm_evidence}
\end{figure}

We conducted two sets of experiments to understand the impact of input size on collective communication efficiency. Results are shown in Figure~\ref{fig:comm_evidence}, which helped identify two ingredients for efficiencies:
\begin{enumerate}
    \item \textbf{Even Input Size:} The Nvidia NCCL~\cite{nccl} library offers efficient collective implementations for all-gather and reduce-scatter that require \emph{even} input tensor sizes across ranks.
    \item \textbf{Larger Input Size:} For fixed communication volume, batching data and issuing fewer collectives improves performance by avoiding the collectives' launch overhead and increasing network bandwidth utilization.
\end{enumerate}

For (1), NCCL's \code{AllGather} API requires even input tensor size and writes outputs into one single tensor. PyTorch's \code{ProcessGroup} wraps the NCCL API and enhances it by supporting uneven input tensor sizes across ranks and allowing users to provide a list of output tensors. The flexibility comes with an efficiency trade-off, as shown in Figure~\ref{fig:comm_evidence}~(a). We use \code{All-Gather Base} to denote NCCL's \code{AllGather} behavior, and \code{All-Gather} to denote the one that takes a list of tensors as outputs. The latter incurs additional copies between the individual output tensors and the consolidated single large output tensor before and after the communication. Moreover, for uneven inputs, \code{ProcessGroup} mimics \code{AllGather}'s behavior using group \code{Broadcast}, which is slower than \code{All-Gather Base}. In the experiments, we created artificial unevenness by moving $1$ element and $1e6$ elements from rank 1 to rank 0 respectively. The results show that the \code{All-Gather Base} with even input size achieved highest efficiency. 

For (2), Figure~\ref{fig:comm_evidence}~(b) fixes the total communication to be $2^{30} \approx 1\text{B}$ \code{FP32} elements and varies the size per \code{All-Gather}, \emph{i.e.}, smaller \code{AllGather} size means more \code{AllGather} invocations. Once the \code{AllGather} size decreases below $33\text{M}$ elements, the total communication time begins increasing rapidly.

\begin{figure}
    \centering
    \includegraphics[width=\linewidth]{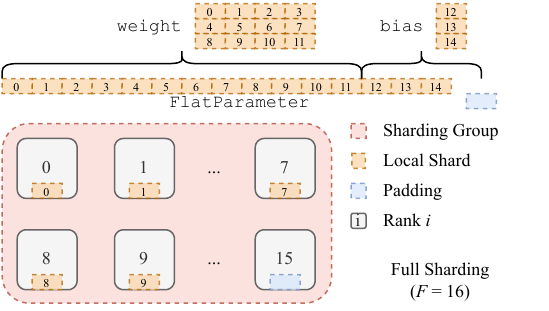}
    \caption{Full Sharding Across 16 GPUs}
    \label{fig:full_sharding}
\end{figure}

Thus, to deliver highly efficient communications, FSDP organizes all parameters within one FSDP unit into a large \code{FlatParameter}, where the \code{FlatParameter} coalesces the communications of its individual parameters and also evenly shards them across ranks. More specifically, the \code{FlatParameter} is a 1D tensor constructed by concatenating $p$ flattened original parameters and padding on the right to achieve a size divisible by the sharding factor. To shard the \code{FlatParameter}, FSDP divides it into equal-sized chunks, where the number of chunks equals the sharding factor, and assigns one chunk per rank. The \code{FlatParameter}'s gradient inherits the same unsharded and sharded shapes from the \code{FlatParameter}, and the \code{FlatParameter} and its gradient own the underlying storage of the original parameters and their gradients, respectively. Figure~\ref{fig:full_sharding} depicts one example, where we use one FSDP unit to shard a $4\times 3$ \code{nn.Linear} layer across 16 GPUs. In this case, every GPU only holds one element from the \code{FlatParameter} with the last rank holding the padded value. 

This flatten-concat-chunk algorithm permits each original parameter to have arbitrary shape while minimizing the required padding (to be at most $F-1$), reflecting its generality. 
%However, the does come at the cost of structure: An original parameter may be split across ranks such that it cannot be reshaped to a sharded version of its original shape. Regardless, 
Moreover, under this algorithm, the sharded and unsharded \code{FlatParameter} and its gradient have the exact data layout expected by \code{AllGather} and \code{ReduceScatter}, respectively. This enables calling the collectives without any additional copies for either the input or output tensors.
% Thus, the \code{FlatParameter} and its gradient represent the atoms for FSDP's collectives.

More formally, suppose for a model with $\Psi$ number of elements, FSDP constructs $N$ \code{FlatParameter}s with numels $\psi_1, \dots, \psi_N$, where $\sum_{i=1}^N \psi = \Psi$. For sharding factor $F$, the peak parameter memory contribution is in $O(\sum_{i=1}^N \frac{\psi_i}{F} + \max_{i=1}^N \psi_i)$ because FSDP always keeps each local sharded \code{FlatParameter} with size $\frac{\psi_i}{F}$ in GPU memory and must materialize each unsharded \code{FlatParameter} with size $\psi_i$ one by one during forward and backward. Since the first $\sum_{i=1}^N \psi_i = \Psi$ is fixed, the peak parameter memory contribution is determined by $\max_{i=1}^N \psi_i$. At the same time, the number of collectives per iteration is in $O(N)$. This evidences FSDP's memory-throughput trade-off: Finer-grained \code{FlatParameter} construction decreases peak memory but may decrease throughput by requiring more collectives. Users can control this trade-off by specifying how to wrap sub-modules into FSDP units.

\subsubsection{Hybrid Sharding}\hfill \break

We refer to the strategy when the sharding factor is greater than $1$ but less than $W$ as \textit{hybrid sharding}, as it combines both sharding and replication. For global world size $W$ and sharding factor $F$, the parameters are sharded within each group $S_1, \dots, S_{W/F}$ and are replicated within each complementary group $R_1, \dots, R_F$, where each $S_i, R_j \subseteq \{1, \dots, W\}$ gives the ranks in the sharded or replicated group, respectively.

\begin{figure}
    \centering
    \includegraphics[width=\linewidth]{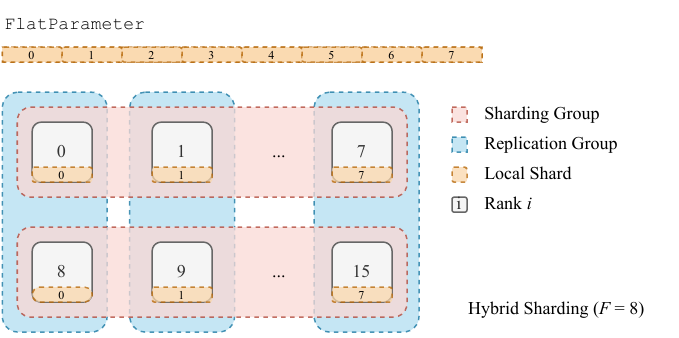}
    \caption{Hybrid Sharding on 16 GPUs: GPUs are configured into 2 sharding groups and 8 replication groups}
    \label{fig:hybrid_sharding}
\end{figure}

For gradient reduction, the single reduce-scatter over all ranks becomes a reduce-scatter within each of the sharded groups followed by an all-reduce within each of the replicated groups to reduce the sharded gradients. The equivalence follows from the decomposition
\begin{equation}
    \sum_{r=1}^W g_r = \sum_{i=1}^{W/F} \sum_{r \in S_i} g_r,
\end{equation}
where $g_r$ represents the gradient on rank $r$.

% \todo[inline]{Figure showing equivalence of global RS and RS + AR}

Hybrid sharding can take advantage of datacenter locality for accelerated training and can reduce cross host traffic to avoid as much contention in the oversubscribed environment as possible. At the same time, it provides a graduating trade-off between memory saving and throughput degradation, which is particularly helpful for models whose required memory footprint when trained with full replication is just slightly above the device capacity and do not want full sharding. Figure~\ref{fig:hybrid_sharding} shows one example. 

Specifically, datacenters typically adopt a fat-tree network topology~\cite{liu2017incbricks} with over-subscription, leading to abundant locality to exploit and a well-motivated reason to reduce cross-host traffic~\cite{luo2020plink}. Hybrid sharding can provide a natural mechanism to map the device mesh into the datacenter layout to exploit such locality. For example, consider a cluster as a group of $W$ accelerators grouped into hosts of of $G$ accelerators each (where the communication among accelerators on the same host is much faster than the communication across hosts), we can set $F = \frac{W}{G}$ to limit the \code{AllGather} (and \code{ReduceScatter}) operations within the same host, while creating a replication group for accelerators with the same local rank across hosts. For an $M$-sized model, we can then compute the total cross-host traffic per GPU in the hybrid setup to be $2M\frac{W-1}{GW}$, a drastic reduction compared to full replication's $2M\frac{W-1}{W}$ and full sharding's $3M\frac{W-1}{W}$. Additionally, since the \code{AllReduce} collectives used in hybrid sharding operates at a smaller world size, they empirically achieve a better performance than invoking collectives at the global scale (in the case of full replication and full sharding), due to straggler effects and larger network interference. 

Another important design motivation for hybrid sharding is the needs from medium-sized models. These models are large enough to cause out of memory issues when trained with full replication but are not large enough to fully utilize accelerator memory when used with full sharding, leading to \emph{both} runtime overhead and memory waste. The hybrid sharding strategy creates a much richer memory-throughput trade-off space by simply adjusting $F$.

% We highlight one particular case of hybrid sharding that is performant for medium-sized models. For context, communication backbones in modern datacenter components are both hierarchical and heterogeneous. The fat-tree topology~\cite{liu2017incbricks} with different link speeds at different levels and over-subscription naturally lead to abundant locality to exploit. Failure to tap into this property often results in subpar scalability and performance~\cite{luo2020plink}. In particular, clusters are typically organized as a group of nodes with each node consisting of a group of accelerators, where the communication is faster intra-node and slower intra-node. Thus, to exploit this heterogeneity, a performant choice of sharding factor is the intra-node size (\textit{e.g.}\ 8 for typical Nvidia GPU nodes), meaning that parameters are only sharded within each node. This changes the global reduce-scatter to an intra-node reduce-scatter followed by an inter-node all-reduce.

\subsubsection{Autograd}
\label{subsubsection:autograd}\hfill \break

FSDP's \code{FlatParameter} must inter-operate with PyTorch's autograd engine to ensure (1) correct gradient propagation and (2) timely gradient reduction. For (1), recall that the \code{FlatParameter} and its gradient own the underlying storage of the original parameters and their gradients, respectively. To achieve this, before forward computation, FSDP sets the original parameters to be views into their unsharded \code{FlatParameter} using \textit{autograd-visible} \code{torch.split()} and \code{torch.view()} calls. Then, the autograd engine naturally allocates the unsharded \code{FlatParameter} gradient and writes each original parameter's gradient to the appropriate offset as defined by \code{torch.split()}'s backward function. For (2), FSDP registers a gradient hook that only runs once the \code{FlatParameter}'s gradient is finalized. The hook represents the post-backward logic and includes the gradient reduction. Notably, FSDP's approach builds on top of PyTorch's autograd engine instead of hacking around it. As a result, FSDP automatically handles unconventional cases such as when not all parameters are used in the forward or when there are multiple forwards before a backward.

\subsection{Communication Optimizations}

The FSDP framework incorporates a range of native communication optimization techniques. This section unveils four major ones: overlapping, backward pre-fetching, forward pre-fetching, and accumulation. 

\subsubsection{Overlapping Communication and Computation} \hfill \break

\begin{figure}
    \centering
    \includegraphics[width=\linewidth]{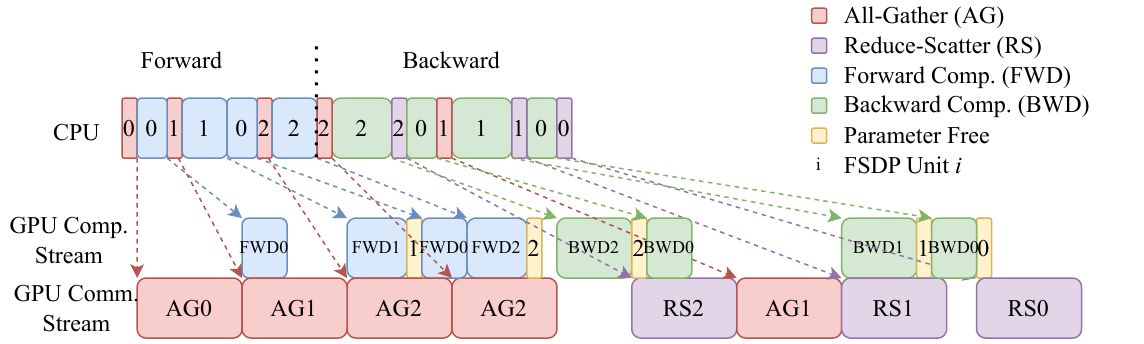}
    \caption{Overlap Communication and Computation}
    \label{fig:overlap}
\end{figure}

The PyTorch \code{c10d} library has a \code{ProcessGroup} abstraction that represents a group of processes that can run collectives together. For the NCCL backend, the \code{ProcessGroupNCCL} implementation has an internal NCCL stream per device, where the separate internal stream is for asynchronous execution with the current stream, which is typically the default stream running computation. Those asynchronous collectives return \code{Work} objects, where calling \code{Work.wait()} blocks the CPU thread until the collective finishes. For general correctness, \code{ProcessGroupNCCL} synchronizes the internal stream with the current stream before running the collective. \code{DistributedDataParallel} leverages the async-collective-and-\code{wait()} approach to overlap the gradient \code{All-Reduce}s with backward computation. However, in contrast to DDP's backward where the \code{AllReduce} \textit{proceeds} the computation with which to overlap, FSDP's forward issues the \code{AllGather} \textit{following} the computation with which to overlap since in eager execution, FSDP cannot know which \code{FlatParameter} to \code{AllGather} next to reorder it before the computation. This difference in kernel-issue order makes following the async-collective-and-\code{wait()} approach infeasible for FSDP. Namely, since \code{ProcessGroupNCCL} synchronizes with the current (default) stream, the \code{All-Gather} will not run until the computation with which to overlap finishes. To address this, FSDP uses a separate CUDA stream to issue the \code{AllGather}s, bypassing the false dependency on preceding computation in the default stream and allowing each \code{AllGather} to overlap. As a result, FSDP's collective synchronization operates on streams, not simply \code{Work} objects. Figure~\ref{fig:overlap} illustrates one example. Note that the backward pass excludes the \code{AG0} \code{All-Gather} because FSDP intentionally keeps the outermost FSDP unit's parameters in memory to avoid redundantly freeing at the end of forward and then re-\code{All-Gather}ing to begin backward.

% As a side optimization, both the \code{ReduceScatter} and \code{AllReduce} (when hybrid shard is on) can run asynchronously with respect to the backward pass, as they are not needed until the optimizer step.

\subsubsection{Backward Prefetching} \hfill \break

FSDP enforces a single CUDA device per rank and uses a single process group for both \code{AllGather} and \code{ReduceScatter}, which means that its collectives run sequentially in the process group's internal NCCL stream. In the backward pass, FSDP issues the \code{ReduceScatter} for the current \code{FlatParameter} and then the \code{AllGather} for the next \code{FlatParameter}. Hence, the single NCCL stream forces the \code{ReduceScatter} to block the next \code{AllGather}, which in turn blocks the next gradient computation and may become exposed on the critical path.

To avoid two consecutive exposed communication calls in the backward pass, FSDP \textit{backward prefetching} issues the next \code{AllGather} before the current \code{ReduceScatter}. However, as mentioned before, a challenge for eager execution is knowing which \code{FlatParameter} to \code{AllGather} next. FSDP resolved this challenge by recording the reverse forward execution order of modules as the proxy of their backward execution order. Moreover, the forward order is freshly recorded each iteration, meaning that the \textit{backward prefetching} is compatible with dynamism across iterations.

%\todo[inline]{Backward prefetching figure: reorder RS and AG}

\subsubsection{Forward Prefetching} \hfill \break

For some workloads with relatively slow CPU execution, the CPU thread may not be able to issue the next forward \code{AllGather} early enough to efficiently fill the NCCL stream. If the model follows a static computational graph across iterations, then FSDP can assume the forward execution order of modules from the previous iteration and prefetch the next \code{AllGather} explicitly in the forward pass. This \textit{forward prefetching} issues the next \code{AllGather} before forward computation of current FSDP unit.

\subsubsection{Gradient Accumulation} \hfill \break

FSDP offers two variations of gradient accumulation: with and without communication. With communication, FSDP still reduces gradients across ranks, and each rank saves the sharded gradients. Simply running multiple iterations without clearing gradients achieves this. Without communication, FSDP does not reduce gradients across ranks, and each rank saves the unsharded gradients. This latter variation trades off increased memory usage with decreased communication, which can increase end-to-end throughput.

\subsection{Memory Management}\label{sec:memory_management}

%\subsubsection{Rate Limiter} \hfill \break
PyTorch uses a CUDA caching allocator as a middle layer to serve GPU allocation and free requests for PyTorch programs. In order to effectively manage memory, FSDP uses a \textit{rate limiter} to take into account the memory impact of the caching allocator on programs that use several CUDA streams and run fast CPU threads.

\subsubsection{How Does PyTorch Caching Allocator Affect Memory}
\label{subsubsection:pytorch caching allocator}\hfill \break

The caching allocator avoids frequent calls to \code{cudaMalloc} and \code{cudaFree}, where the latter incurs a costly device synchronization. Specifically, the caching allocator requests CUDA memory blocks and internally determines how to split and reuse the blocks without returning them to CUDA with the goal being to reach a steady state without further calls to \code{cudaMalloc} and \code{cudaFree}.

The caching allocator runs from the CPU thread, meaning that it must decide which caching allocator block to use for an allocation when the \textit{CPU thread} processes the allocation request. It cannot wait until the GPU kernel needing the allocation actually runs, which may be much later. 

For a single stream, the caching allocator can directly reuse memory blocks by the stream's sequential ordering semantics. However, for separate producer and consumer streams, there are no inter-stream ordering guarantees, and the caching allocator cannot be certain that a block is safe to reuse until the last \textit{GPU kernel} depending on that memory finishes running. Hence, if the CPU thread runs far ahead of the GPU execution, then the caching allocator cannot reuse blocks for the producer stream with pending GPU kernels from the consumer stream.

Furthermore, caching allocator blocks are allocated \textit{per stream} and cannot be reused for a different stream, this over-allocates blocks to the producer stream that could otherwise be used for the consumer stream (\textit{e.g.}\ for activations). The GPU itself may have enough memory to serve a new allocation in the consumer stream, but the overallocation to the producer stream may lead to the caching allocator failing to serve it. This forces a blocking sequence of \code{cudaFree}s to reset the caching allocator memory state called a \code{cudaMalloc} retry that greatly degrades training throughput.

\subsubsection{Rate Limiter}
\label{subsubsection:rate limiter}\hfill \break

FSDP allocates the \code{AllGather} destination tensor representing the unsharded \code{FlatParameter} in a producer stream, and the forward and backward computations using the \code{AllGather}ed parameters run in a consumer stream (typically the default stream). For a fast CPU thread, there may be pending GPU computation kernels when the caching allocator must serve the next \code{AllGather}, leading to no block reuse. Even after the blocks are not active in the \code{AllGather} producer stream, these reserved blocks can not serve default computation stream's allocation requests, and thus may force blocking \code{cudaFree}s and \code{cudaMalloc}s. 

%\todo[inline]{Add figures showing that CPU allocation must be served after previous GPU computation to reuse the block}

FSDP offers a \textit{rate limiter} that intentionally blocks the CPU thread to ensure proper caching allocator block reuse. It allows at most two inflight \code{AllGather}s, which is the minimum amount to still achieve communication and computation overlap.

\section{Implementation}\label{sec:implementation}

This section delves into the intricacies of FSDP implementation, which although do not alter the FSDP core algorithm, are crucial to understand before adopting FSDP. 

%\subsection{Program Interface}

Users can access FSDP through two APIs, \code{FullyShardedDataParallel} model wrapper and \code{fully\_shard} module annotator. The former wraps the entire model and replaces sub-modules with corresponding FSDP units. In contrast, the latter installs FSDP logic as \code{nn.Module} forward and backward hooks, preserving both model structures and parameter fully-qualified names. 
\subsection{Initialization}

Section~\ref{sec:full_shard} described FSDP's solution to efficiently initialize large models, which works well when sub-module initializations are self-contained. In a rare situation where one sub-module's initialization depends on a parameter from the different sub-module, the on-demand materialization and record-replay approach might break if the parameter belongs to a different FSDP unit, because the unsharded version of that parameter could have been discarded to reduce memory footprint. Therefore, besides the advanced deferred initialization, FSDP offers two more options:

\begin{itemize}
    \item \textbf{Initialize unsharded model on GPU}. The memory requirement for model initialization may be smaller than that for training since training also involves gradients, activations, and optimizer states. Consequently, if the training step cannot be performed on a single GPU device, users might still be able to initialize the entire model on a GPU and pass it to FSDP. Then, optimizers should be instantiated after FSDP shards the model, to reduce the memory footprint and align with the sharded gradients produced by FSDP.
    \item \textbf{Initialize unsharded model on CPU}. If the size of the unsharded model surpasses the capacity of GPU memory and can only be accommodated in CPU memory, it becomes impracticable to move the unsharded model entirely to the GPU before handing it over to FSDP for parameter sharding. To overcome this challenge, FSDP adopts a streaming approach, where the model is migrated to the GPU unit by unit. Upon arrival to the GPU, the parameters of each unit are immediately sharded, which in turn reduces the memory overhead before processing the next unit. This approach remains viable even when there are cross-submodule dependencies during initialization, given that all parameters of the entire unsharded model are present in the CPU memory.
\end{itemize}

%The goal of FSDP's model initialization is to shard the parameters across ranks, where together they represent the original unsharded parameters. As input, FSDP can take models whose sharded size falls into one of three regimes: (1) fitting in GPU memory, (2) fitting only in CPU memory, and (3) not fitting in CPU memory. For each, FSDP offers a model initialization solution.

%\subsubsection{Fitting in GPU Memory}
%Even if the model's unsharded size fits in GPU memory, some form of sharding may be necessary for training due to other memory costs such as from gradients, optimizer states, and activations, in which case FSDP is one option. For this regime, the user can simply initialize the unsharded model on CPU, move it to GPU, and pass it to FSDP for parameter sharding.

%\subsubsection{Fitting Only in CPU Memory}
%If the unsharded model size does not fit in GPU memory and only CPU memory, then the unsharded model cannot be atomically moved to GPU before passing to FSDP for parameter sharding. Instead, FSDP supports a streaming approach, where the model is moved layer by layer to GPU, where each layer's parameters are sharded immediately on GPU, reducing its memory overhead before processing the next layer.

Note that both approaches above are subject to their own limitations. The first method entails the entire model fitting within a single GPU device and thus becomes infeasible for larger models. The second method, on the other hand, can handle larger models since the CPU has considerably larger memory. However, this approach may experience substantial slowdowns in comparison to deferred initialization due to the limited memory bandwidth and parallelization capabilities of the CPU. In light of these observations, users may still prefer deferred initialization, even when dealing with models of the size range encompassed by the previous two methods.

%\subsubsection{Not Fitting in CPU Memory}
%For models that cannot even fit in CPU memory, FSDP supports \textit{deferred initialization}, where the parameters are not materialized until after sharding. This deferred initialization can offer significant speedups since the work done by the memory-bandwidth-bound tensor fill kernels is partitioned across the workers and happens directly on GPU leveraging its high bandwidth memory without ever running on CPU. Given this, users may prefer deferred initialization even if the model size falls under either of the two preceding regimes.

%\todo[inline]{Andrew: our deferred init not in great shape (parity with unsharded randomness)}

To delimit the scope of each FSDP unit, users may choose to employ the \code{FullyShardedDataParallel} wrapper by intrusively applying it to sub-modules in model source code, or alternatively, provide a custom function to the \code{auto\_wrap\_policy} argument upon instantiation. Selecting the optimal wrapping approach typically requires some experiments and measurements.

\subsection{Flat Parameters}

The \code{FlatParameter} class inherits from \code{nn.Parameter} and behaves like an \code{nn.Parameter}. FSDP implements an accompanying \code{FlatParamHandle} class that is responsible for managing individual \code{FlatParameter} instances. The frontend, either \code{FullyShardedDataParallel} or \code{fully\_shard}, interfaces with the \code{FlatParameter}s only through \code{FlatParamHandle}.

One \code{FlatParameter} accommodates storage for all parameter tensors within one FSDP unit. The boundary of the FSDP unit controls the timing for \code{AllGather} and \code{ReduceScatter}, which has a direct impact on overall FSDP performance. In the ideal case, FSDP unit boundaries should align with model execution order. 

%\todo[inline]{assign to Shen, need to explain tensor storage somewhere, maybe in background}

FSDP has access to the model's static \code{nn.Module} structure at construction time. Fortunately, although this structure does not guarantee to faithfully represent model execution order, model authors conventionally translate layers and broader blocks to nested \code{nn.Module} definitions that may naturally have the desired parameter locality. FSDP can leverage that structure to choose the \code{FlatParameter} construction. Indeed, FSDP supports annotating \code{nn.Module}s and follows a simple rule: All parameters in the annotated \code{nn.Module} are assigned to one \code{FlatParameter}, excluding those parameters already assigned. This rule lends itself naturally to nested annotation, where blocks are annotated, forming well-sized \code{FlatParameter}s, and any residual parameters are assigned to their parent. 

Another approach we explored is using the execution order and reconstructing \code{FlatParameter}s dynamically. This approach starts with an initial small \code{FlatParameter} construction, runs a possibly inefficient first iteration while observing the execution order, and reconstructs the \code{FlatParameter}s by coalescing the existing small \code{FlatParameter}s according to the observed order.
\subsection{Runtime}

FSDP augments a local model instance by incorporating communication operations to reduce gradients and gather parameters. Timely initiation of these operations is of paramount importance for ensuring both correctness and efficiency. Starting communication too soon would cause the parameters or gradients with pending updates to be consumed, while initiating communication too late would result in wasting network bandwidth and delay in subsequent computations.

To insert communication-related code to the model forward pass, the \code{FullyShardedDataParallel} \code{nn.Module} wrapper overrides \code{nn.Module}'s \code{forward()} method to install pre-forward and post-forward logic, whereas the functional \code{fully\_shard} implements them by registering \code{nn.Module} hooks through methods such as \code{register\_forward\_pre\_hook()} and \code{register\_forward\_hook()}. It is more challenging to capture appropriate signals from the backward pass, as PyTorch automatically and transparently handles the backward pass. Fortunately, the autograd engine exposes a variety of hooks that enable the installation of custom logic with precise granularity.

\begin{itemize}
    \item \textbf{Hooks on \code{Tensor}} through \code{register\_hook()} allows to run custom function when the gradient of a \code{Tensor} is generated. This can help anchor FSDP logic to an activation's gradient computation in the backward pass. FSDP registers this type of hook to the forward output tensor of every FSDP unit to insert communications before backward pass enters that FSDP unit. 
    \item \textbf{Hooks on \code{backward()}} through \code{queue\_callback()} run right before exiting the current autograd \code{GraphTask}, which is usually the end of overall backward pass. FSDP relies on this hook to wait for pending communications so that the subsequent optimizer step will not consume gradients too early.
    \item \textbf{Hooks on \code{AccumulateGrad}} autograd function fires when the gradient of a parameter has finished accumulation in the current backward pass. FSDP attaches this type of hook to each \code{FlatParameter}'s \code{AccumulateGrad} function to immediately launch \code{ReduceScatter} when gradients are ready. Note that the \code{Tensor} hook mentioned above can potentially achieve the same behavior, but might incur unnecessary delay as it needs to wait for gradient computations for input activations as well. 
\end{itemize}

The aforementioned methodologies collectively integrate the FSDP algorithm with the PyTorch \code{nn.Module} and autograd engine in a non-intrusive and efficient manner.

%The runtime implementation mirrors the description from \ref{sec:sharding_strategy}. Each \code{FlatParamHandle} implements \code{unshard} and \code{reshard} operations, and FSDP calls them from pre-forward, post-forward, pre-backward, and post-backward functions.

% FSDP implements the pre-backward and post-backward logic as hooks. The pre-backward hook is registered on the output tensors that require gradients from the corresponding module's forward pass and is ensured to only run once per backward. The post-backward hook is registered on 

%\todo[inline]{Discuss backward in context of autograd and hooks}
%\todo[inline]{Discuss forward as either module override or hooks}

%FSDP implements the pre-backward logic by registering a hook on the corresponding module's forward output tensors that runs before the gradient computation with respect to the tensor, where the hooks are only registered on the tensors that require gradients since their existence implies the need for gradient computation using the module's parameters and hence the need for a pre-backward \code{unshard} operation. FSDP implements the post-backward logic by registering a hook on the \code{FlatParameter}'s \code{AccumulateGrad} object that runs after all corresponding gradients have been computed (as discussed in \ref{subsubsection:autograd}).

\subsection{Native Mixed Precision}
FSDP offers a versatile native mixed precision mechanism. In terms of parameter management, it adheres to the standard mixed precision technique, which maintains both low and full precision copies of parameters~\cite{micikevicius2017mixedprecision}. Forward and backward computation use the low precision, and the optimizer step uses full precision. FSDP permits user-specified precisions for parameters, gradient reduction, and non-trainable buffers, each independently if desired.

For $\Psi$ number of parameter elements (\code{torch.numel}), $K_{\text{low}}$ bytes per low precision element, and $K_{\text{full}}$ bytes per full precision element, this approach to mixed precision normally increases the memory overhead from $K_{\text{full}}\Psi$ to $(K_{\text{low}} + K_{\text{full}})\Psi$ due to maintaining both precision copies. However, FSDP can sidestep the problem given our design to always keep each local sharded \code{FlatParameter} in GPU memory and only dynamically allocate the unsharded \code{FlatParameter}. For $N$ \code{FlatParameter}s with numels given by $\psi_1, \dots, \psi_N$, the parameter peak memory contribution for FSDP actually \textit{decreases} from $\frac{K_{\text{full}}}{F}\sum_{i=1}^N \psi_i + K_{\text{full}} \max_{i=1}^N \psi_i$ to $\frac{K_{\text{full}}}{F}\sum_{i=1}^N \psi_i + K_{\text{low}} \max_{i=1}^N \psi_i$ bytes. In other words, FSDP directly reduces the second $K_{\text{full}} \max_{i=1}^N \psi_i$ term to $K_{\text{low}} \max_{i=1}^N \psi_i$.

In contrast to \code{torch.amp.autocast} that performs just-in-time casts at the operator level, FSDP's native mixed precision only incurs a full-to-low-precision cast per \code{FlatParameter} in its pre-forward and, if resharding after forward, its pre-backward. Moreover, FSDP's mixed precision permits running all collectives in the low precision, which saves communication volume.

% \todo[inline]{Quantitatively compare autocast vs FSDP MP?}

Users most commonly choose \code{FP16} or \code{BF16} as the low precision and \code{FP32} as the full precision. \code{FP16}'s smaller dynamic range compared that of \code{FP32} exposes \code{FP16} to greater risk of numeric underflow and overflow. The standard solution includes a gradient scaler~\cite{AMPGradScaler} that scales gradients to a safe magnitude. However, since FSDP shards gradients across ranks, a normal local gradient scaler implementation breaks mathematical equivalence, and instead, FSDP provides its own sharded gradient scaler.

%\todo[inline]{Execution order}

% To-do: Should we include code snippet showing torch.cat(param.flatten() for param in params, dim=0)?

% To-do: The FlatParameter has unsharded and sharded states.

%\subsection{\code{FlatParameter} Construction}
%FSDP offers two ways to construct \code{FlatParameter}s: by manual annotation or programmatically  with a policy.

% To-do: manual wrapping; auto wrapping

\section{Evaluation}\label{sec:evaluation}

We conducted an empirical evaluation of FSDP on large language models and recommendation system models and compared the results with those of DDP. Experiment specifications are described in Section~\ref{sec:exp_setup}. Then, we organize experiments into three categories. Section~\ref{sec:eval_efficiency} focuses on how well FSDP handles different sizes of models. Then, Section~\ref{sec:eval_rate_limiting} discusses the impact of throttling communications. Finally, Section~\ref{sec:eval_prod} demonstrate FSDP's ability to scale to gigantic models. 

\begin{figure*}[!htbp]
\begin{minipage}[c]{0.33\textwidth}
  \vspace{0.1em}
  \centering
  \includegraphics[width=\linewidth]{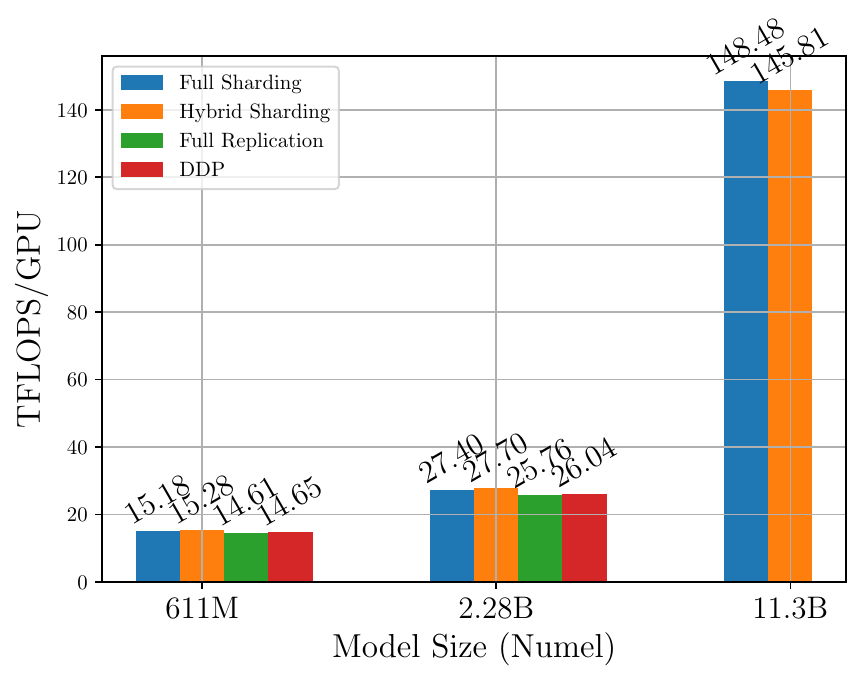}
  (a) Model Scale
\end{minipage}
\begin{minipage}[c]{0.33\textwidth}
  \vspace{1em}
  \centering
  \includegraphics[width=\linewidth]{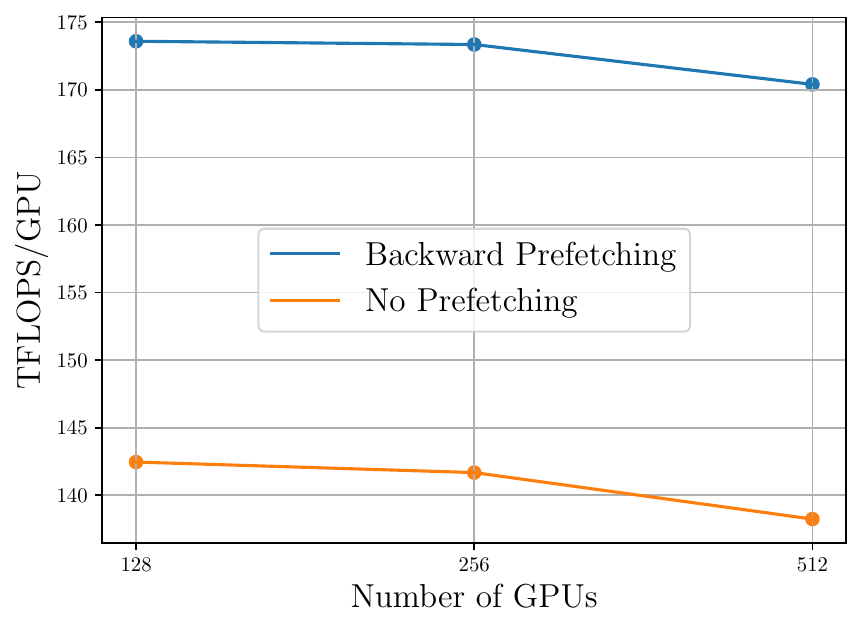}
  (b) GPT-175B Backward Prefetch
\end{minipage}
\begin{minipage}[c]{0.33\textwidth}
  \centering
  \includegraphics[width=\linewidth]{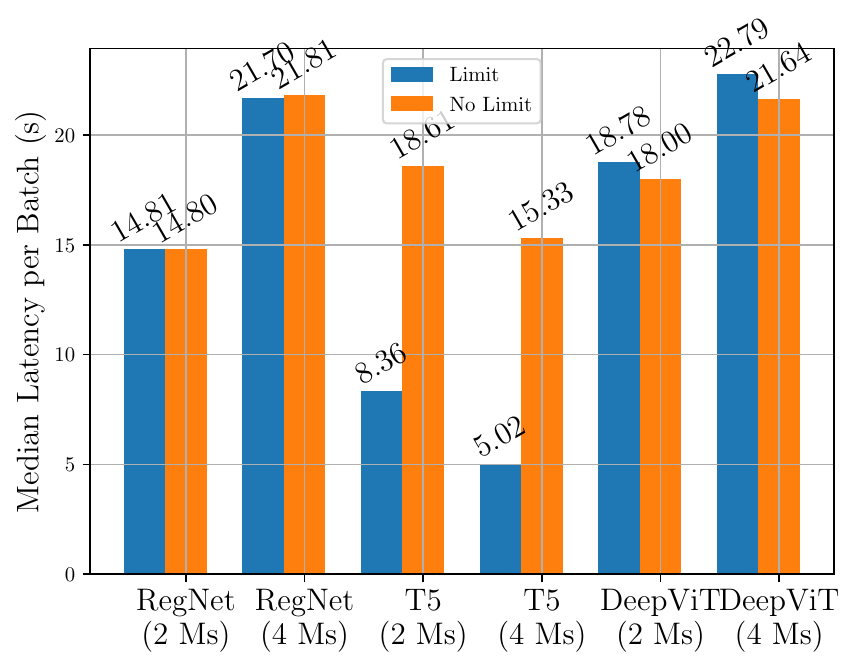}
  (c) Rate Limiter (Ms = Machines)
\end{minipage}
%\vspace{-1em}
\caption{Model Scale and Training Efficiency}\label{fig:mix}
%\vspace{-1em}
\end{figure*}

\subsection{Experiment Setup}\label{sec:exp_setup}

In these experiments, we conducted evaluations on the HuggingFace T5-11B transformer~\cite{raffel2020exploring}, minGPT-175B transformer~\cite{gpt3}, and DHEN recommendation model~\cite{DHEN}. The recommendation models consist of 768B sparse parameters and 550M dense parameters, the sparse parameter tensors were sharded using the first approach mentioned in Section 2.3, which communicates activations instead of parameters, while the dense parameters were trained using FSDP on 8 to 512 A100 80GB GPUs interconnected by a 2Tb/s RoCE network. The objective was to assess the capability and scalability of FSDP in training large-scale models. Additionally, we employed T5-611M, T5-2B and T5-11B transformers to evaluate the performance of various sharding strategies, communication efficiency of prefetching, and communication throttling using rate limiter. Metrics employed in these experiments included TFLOPS per GPU, latency per batch, peak memory allocated, peak memory active, and peak memory reserved.

\subsection{Model Scale}\label{sec:eval_efficiency}

In this section, we investigate the performance of FSDP when dealing with models of different sizes, spanning from 611M to 175B, and make a comparison with DDP~\cite{ddp}.

%\begin{figure}
%    \centering
%    \includegraphics[width=\linewidth]{figure/prefetching_tflops.pdf}
%    \caption{Backward Prefetch On/Off: with 80GB A100, 8 GPUs per Machine}
%    \label{fig:prefetching_tflops}
%\end{figure}

The experimental results on T5 models are displayed in Figure~\ref{fig:mix}~(a). The performance of FSDP and DDP is similar when evaluating 611M and 2.28B models. However, DDP encounters an out-of-memory error when attempting to wrap models larger than 2.28B. In contrast, FSDP can effortlessly accommodate the 11B model and achieve significantly higher TFLOPS by turning on \code{BF16}. These experiments illustrate that practitioners can utilize FSDP for both small and large models, and seamlessly transition across different model configurations.

Then, we conduct additional experiments to measure the acceleration attained through backward pre-fetching. This time we use a larger GPT-175B model, where communication overhead is more prominent. Results are presented in Figure~\ref{fig:mix}~(b), where pre-fetching leads to approximately 18\% speedup, and this TFLOPS gain persists across different GPU cluster sizes. Therefore, for subsequent experiments, we always turn-on backward pre-fetching.

\subsection{Throttle Communications}\label{sec:eval_rate_limiting}

In the subsequent analysis, we investigate the implications of throttling FSDP communications. As expounded in Section~\ref{sec:memory_management}, launching \code{AllGather} too aggressively can lead to unnecessarily high memory footprint, as the CPU thread needs to allocate CUDA memory blocks when the communication kernel is added into the CUDA stream. This predicament may sometimes result in significant performance problems when the CPU thread runs too fast in comparison to CUDA streams. To gauge its efficacy in varying scenarios, we apply rate limiting to three different types of models and applied the maximum feasible batch size in each experiment.

\begin{itemize}
    \item \textbf{RegNet}~\cite{schneider2017regnet}: model size 9B, and batch size 48 for 2 nodes and 72 for 4 nodes.
    \item \textbf{T5}~\cite{raffel2020exploring}: model size 11B, and batch size 2. 
    \item \textbf{DeepViT}~\cite{zhou2021deepvit}: model size 8B, and batch size 105 for 2 nodes and 120 for 4 nodes. 
\end{itemize}

%\begin{figure}
%    \centering
%    \includegraphics[width=\linewidth]{figure/rate_limiter_latency.pdf}
%    \caption{Rate Limiter On/Off: with 40GB A100, 8 GPUs per Machine %    \label{fig:rate_limiter_latency}
%\end{figure}

Experiment results are plotted in Figure~\ref{fig:mix}~(c).  One notable observation is that the rate limiter's effectiveness is not consistent, as it does not attain any speedups in the RegNet experiments, and even impedes the DeepViT ones. This behavior is expected since throttling the communications can only boost training if the fast CPU thread aggressively allocates GPU memory blocks and causes defragmentations. If it is difficult to identify with certainty from latency measurements or profiled traces, CUDA \code{malloc} retry can serve as a helpful indicator, which can be obtained from the \code{num\_alloc\_retries} key in the \code{torch.cuda.memory\_stats()} dictionary.

The experiments conducted with T5 models have demonstrated that the rate limiter technique can greatly benefit training efficiency, yielding up to 5X speedups. However, for DeepViT models, introducing communication throttling can result in an additional 5\% overhead. This is due to the fact that delaying the \code{AllGather} communication can potentially block subsequent model computations that rely on the \code{AllGather}ed parameters, especially in cases where communication is the dominant factor. Therefore, before enabling rate limiting, practioners should verify whether defragmentation has taken place during training.

\begin{figure*}[!htb]
\begin{minipage}[c]{0.33\textwidth}
  \centering
  \includegraphics[width=\linewidth]{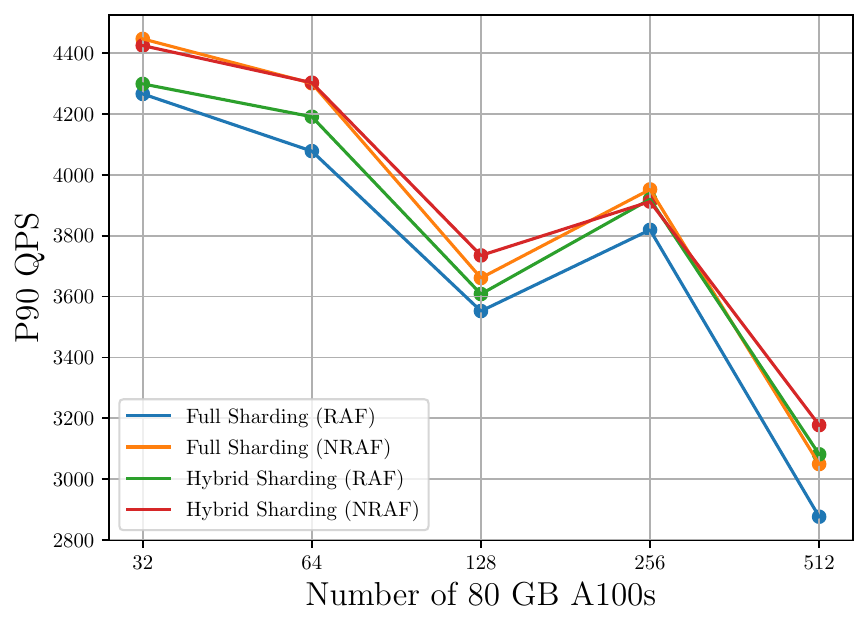}
  (a) DHEN QPS
\end{minipage}
\begin{minipage}[c]{0.33\textwidth}
  \centering
  \includegraphics[width=\linewidth]{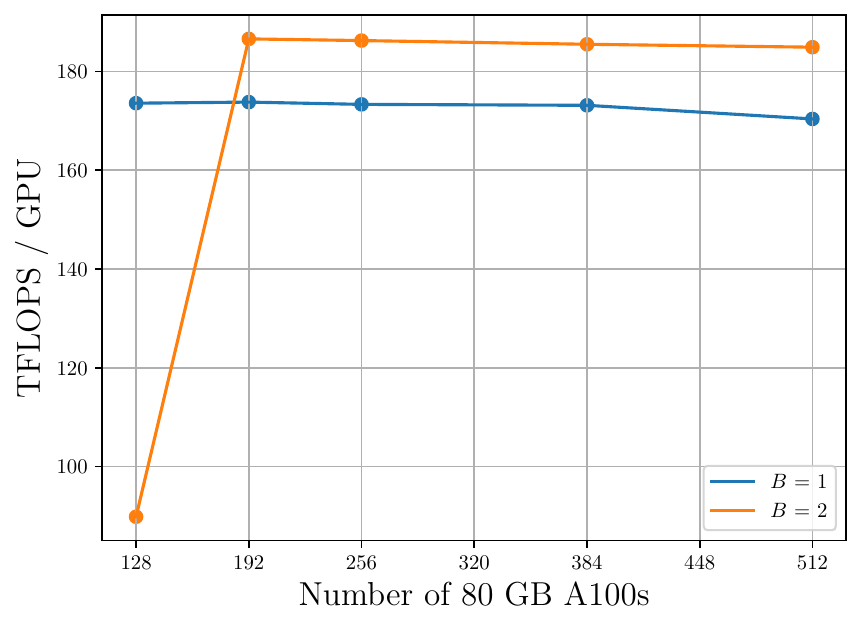}
  (b) GPT-175B TFLOPS
\end{minipage}
\begin{minipage}[c]{0.33\textwidth}
  \centering
  \includegraphics[width=\linewidth]{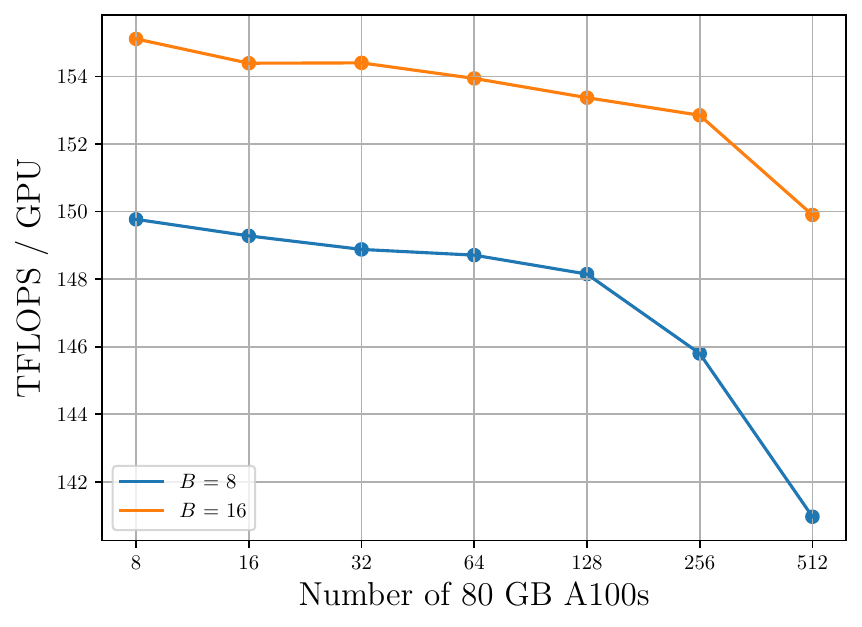}
  (c) T5-11B TFLOPS
\end{minipage}
%\vspace{-1em}
\caption{Training Throughput: To conform with DHEN convention, we use sample/ GPU/second (QPS) for DHEN.}\label{fig:throughput}
%\vspace{-1em}
\end{figure*}

\begin{figure*}[!htb]
\begin{minipage}[c]{0.33\textwidth}
  \centering
  \includegraphics[width=\linewidth]{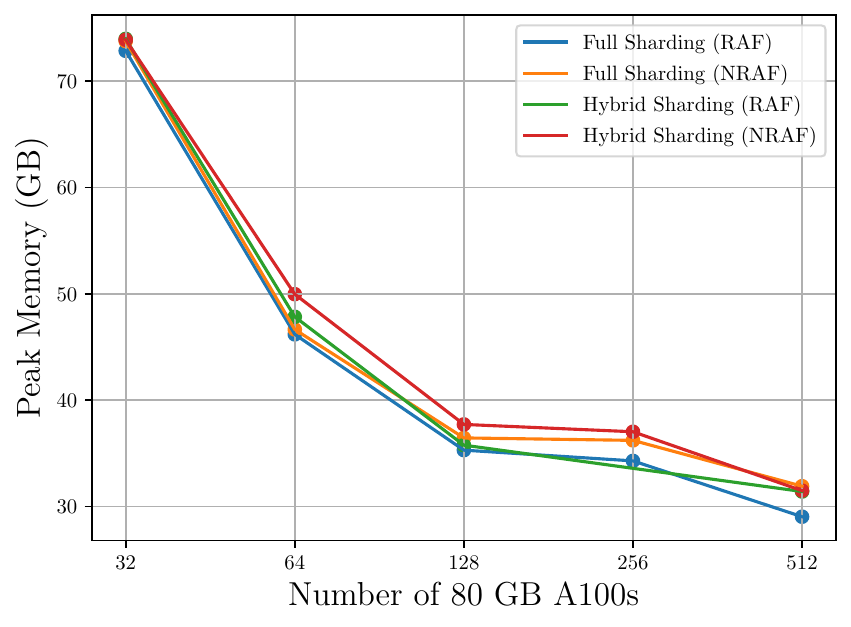}
  (a) DHEN
\end{minipage}
\begin{minipage}[c]{0.33\textwidth}
  \centering
  \includegraphics[width=\linewidth]{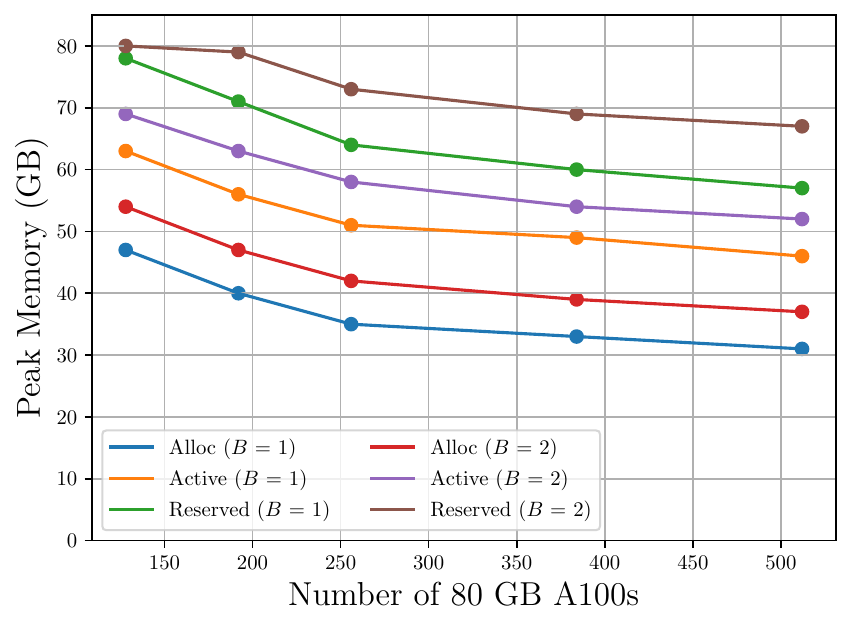}
  (b) GPT-175B
\end{minipage}
\begin{minipage}[c]{0.33\textwidth}
  \centering
  \includegraphics[width=\linewidth]{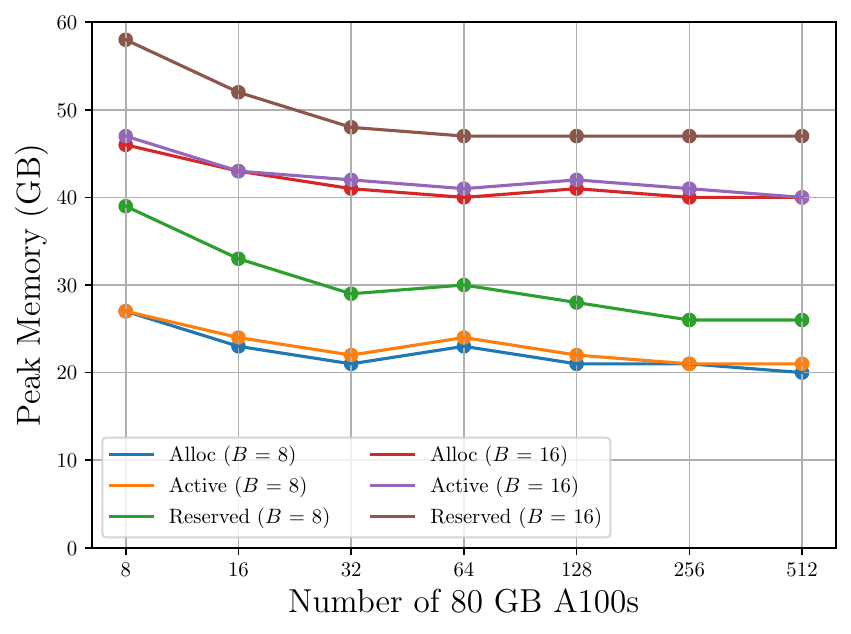}
  (c) T5-11B
\end{minipage}
%\vspace{-1em}
\caption{Memory Footprint}\label{fig:memory}
%\vspace{-1em}
\end{figure*}

\subsection{Efficient Training for Large Models}\label{sec:eval_prod}

To evaluate capability of using FSDP for large models, we ran three types of models using Full Sharding with prefetching and rate limiter turned on. Activation checkpointing and \code{BF16} mixed precision are also applied in these experiments. Adam optimizer is used to reflect a production workload setup and to incur the costly two optimizer states per parameter.
\begin{itemize}
    \item \textbf{DHEN large recommendation model}~\cite{DHEN}: model size - 768B sparse parameters and 550M dense parameters, and batch size 1024. 
    \item \textbf{minGPT transformer}~\cite{MinGPT}: model size 175B, vocab size 50000, block size 2048, batch size 1 and 2 for 128, 192, 256, 384 and 512 GPUs.
    \item \textbf{HuggingFace T5 transformer}~\cite{raffel2020exploring}: model size 11B, sequence length 512, batch size 8 and 16 for 8, 16, 32, 64, 128, 256, 512 GPUs.
\end{itemize}

In the DHEN experiments, we further combine sharding strategies with two different configurations: 

\begin{itemize}
    \item \textbf{RAF:} reshard-after-forward frees \code{AllGather}ed shards from other GPUs after forward pass and unshards them again before backward computation. This reduces peak memory consumption at the cost of higher communication overhead.
    \item \textbf{NRAF:} no-reshard-after-forward is the opposite where the unsharded model parameters stay in GPU memory after forward pass until backward computations finish, which trades higher memory footprint for lower communication overhead.
\end{itemize}

The experimental results in Figure~\ref{fig:throughput}~(a) and Figure~\ref{fig:memory}~(a) indicate that FSDP is capable of accommodating DHEN models on a large GPU cluster. It was observed that Full Sharding with RAF yields the smallest memory footprint but with a corresponding trade-off of reduced QPS. Conversely, Hybrid Sharding with NRAF demonstrated the opposite behavior, as it has employs both a smaller sharding group and skips one reshard. When adding more GPUs to in the cluster, the peak memory usage consistently decreases as a result of a decrease in the size of each rank's model shard.

With the 175B model, the experiments achieved more than 173 and 186 TFLOPS per GPU with batch size equal to 1 and 2 respectively as shown in Figure~\ref{fig:throughput}~(b). This is equivalent to approximately 55\% and 60\% of GPU hardware utilization, given that the A100's peak is 312 TFLOPS using the \code{BF16} tensor core. Furthermore, the model demonstrated linear scalability from 128 GPUs to 512 GPUs, in terms of TFLOPS, which affirms the efficacy of FSDP in handling large models with expensive computations or high-speed network interconnections. Notably, with 128 GPUs, setting the batch size to 2 resulted in a considerably lower per-GPU TFLOPs in comparison to other scenarios. This was due to CUDA memory defragmentation during the backward pass. The backward pass contributed 85.56\% of the iteration latency for the 128 GPU batch size equals 2 case, while a normal backward pass only accounted for about 67\% in these experiments. Using 128 GPUs is more likely to trigger defragmentation, as each GPU needs to accommodate a larger model shard. Figure~\ref{fig:memory} confirms this explanation, where the PyTorch CUDA caching allocator depletes all 80GB of the CUDA memory as shown on the top left corner.

Finally, for T5-11B models as shown in Figure~\ref{fig:memory}~(c), all experiments are executed comfortably below GPU memory capacity, where defragmentations are unlikely to happen. Nevertheless, as the number of GPUs increases from 8 to 512, a 7\% regression in per-GPU TFLOPS is still evident as illustrated in Figure~\ref{fig:throughput}~(c). This suggests that communications begin to outweigh computations on large clusters, and a near-perfect overlap between communication and computation is no longer attainable.
\section{Related Work}\label{sec:related}

The DDP~\cite{ddp} model wrapper, which is based on the model replication design, was an initial distributed training feature introduced in PyTorch~\cite{pytorch:nips:2019}. Although it can handle large datasets, it cannot accommodate the ever-increasing model sizes that are now prevalent in the field.

ZeRO~\cite{zero, zero:offload} and cross-replica sharding~\cite{xu2020automatic} inspired the FSDP design, but FSDP is intrinsically different. Prior work employs model partitioning or per-parameter sharding to distribute parameter tensors, and rely on \code{Broadcast} and \code{Gather} collective communication primitives to synchronize values. Although this design can achieve the same functionality, it could lead to uneven workload distribution across GPU devices, which hampers the efficiency of synchronized distributed training. Additionally, since this approach modifies the internals of the machine learning framework, such as tensor storage and memory management, it might no longer work when the internal implementation is updated or new features are introduced. Therefore, a native solution that is co-designed with the core components of the framework would provide a more robust and consistent user experience.

MiCS~\cite{zhang2022mics} and FSDP differ in gradient communication strategies. MiCS uses a global \code{AllReduce} followed by sharding within each partition group, whereas FSDP employs \code{AllGather} and \code{ReduceScatter}. As a result, each rank in MiCS must hold the entire model gradients, leading to higher memory usage than FSDP's approach of sharding a single layer. While both MiCS and FSDP use a hybrid communication strategy to improve efficiency at scale, FSDP's approach schedules \code{AllGather} within a flexibly-sized sharded group, potentially resulting in lower runtime latency than the two-hop \code{AllGather} utilized by MiCS. This reduced latency is crucial as the \code{AllGather} operation is critical to execution, and limiting the world size and participants of \code{AllGather} to accelerators within a group with good locality can result in lower latency and higher throughput.

Pipeline parallelism~\cite{gpipe, harlap2018pipedream} involves partitioning model parameters and their activations across multiple devices through the division of models into pipeline stages. However, pipeline parallelism requires model changes and meticulous tuning for microbatch sizes, number of stages and partitions, as well as intricate scheduling procedures to optimize performance by squeezing out bubbles. 

Additionally, specific attention is given to high profile architectures such as transformers. For example, sequence parallelism~\cite{sequence} reduces activation memory in conjunction with tensor parallelism; Pipetransformer~\cite{pipetransformer} designed a dynamic 2D parallelism that allows changing the dimensions of pipeline and data parallelism on the fly, depending on learning signals. These methods are highly effective but can be difficult to generalize as they either rely on the specific implementation or the model's layered structure.

Many existing solutions combine data parallelism with other parallelisms to achieve speedup. For example, Megatron~\cite{narayanan2021efficient} demonstrated highly efficient deep transformer training on large clusters using 3D (data, tensor and pipeline) parallelism. Further, compiler-based techniques such as Alpa~\cite{zheng2022alpa}, GSPMD~\cite{gspmd}, and FlexFlow~\cite{flexflow} leverage profiling, performance modeling, user annotations and search to find the best configuration across the parallelism space of data, tensor and pipeline for a given cluster. In all cases, FSDP provides the benefit of being a drop-in replacement for data parallelism that reduces data redundancy along the data parallel axis.

Orthogonal memory-saving techniques include gradient compression~\cite{compression}, mixed-precision training~\cite{precision}, tensor rematerialization~\cite{DTR} and CPU-offloading~\cite{fang2022parallel}, but they could have implications on model accuracy and incur overhead in (un)compression, quantization, recomputation, and host-to-device copies, respectively.
\section{Discussion}\label{sec:discussion}

This section discusses how FSDP can be combined with other parallelism paradigms and known limitations when adopting FSDP. 

\subsection{FSDP Interoperability}

Further increasing scalability and efficiency of distributed training requires combining FSDP with other paradigms. This section briefly highlights how the FSDP design enables mixing and matching with other types of parallelisms.

\subsubsection{Pipeline Parallelism} \hfill \break

Pipeline parallel can be functionally integrated with FSDP by employing FSDP to wrap each individual pipeline stage. However, as pipeline parallel divides input mini-batches into smaller micro-batches, the default full sharding strategy in FSDP would have to unshard model parameters for every micro-batch. Consequently, combining these approaches with default FSDP configurations may lead to significant communication overhead. Fortunately, FSDP offers alternative sharding strategies that can keep parameters unsharded after the forward pass, avoiding unnecessary \code{AllGather} communications per micro-batch. Admittedly, this requires storing parameters of an entire pipeline stage on the GPU device, but FSDP can still reduce memory usage as it still shards gradients and optimizer states.

\subsubsection{Tensor Parallelism} \hfill \break

In contrast to FSDP, tensor parallel keeps parameters sharded during computation, which is necessary if any sub-module is too large to fit in GPU memory. Presently, PyTorch provides a prototype feature called \code{parallelize\_module} that can be combined with FSDP to construct 2D parallelism. It works by organizing devices into a 2D mesh where PyTorch's distributed tensor \code{DTensor} manages tensor parallelism on one dimension and FSDP applies sharded data parallelism on the other dimension. These two dimensions communicate activations and parameters, respectively. We usually keep the tensor-parallel communications, which block subsequent computation, intra-node to leverage the higher network bandwidth, and allow the FSDP communications operate on the other mesh dimension inter-node.

%\todo[inline]{Probably need someone to talk about Micro-batching}

%\subsubsection{3D Parallelism (3DP)}
%Because FSDP can serve as a drop-in replacement for DDP, FSDP's support for 3D Parallelism in the style of Alpa~\cite{zheng2022alpa} or Megatron~\cite{narayanan2021efficient} that applies intra-host TP, cross-host PP and cross-stage (of pipeline) data parallel is straightforward. By wrapping each pipeline stage, FSDP brings additional memory savings as it removes the memory redundancy across pipelines.

%\subsubsection{Hybrid Parallelism}
%Special attention must be given to emerging workloads that require different training paradigms for different components. For example, in modern recommendation models~\cite{DLRM,DHEN}, embedding tables are often used at the beginning of each iteration to convert sparse inputs to dense representations. The sheer sizes of the table and their nature of sparse updates make most of them unfit for data parallelism (including FSDP). Thus, state-of-the-art frameworks employ specialized sharding and synchronization algorithms for the embedding architecture to use in conjunction data parallelism for the dense components~\cite{mudigere2021high}. To support this use case, FSDP adopts an ignore-list approach that excludes certain modules from being synchronized and thus leaves the embedding architecture intact. To reduce interference, the communication for the embedding architecture and the collectives used in FSDP can operate in separate device streams.

\subsection{Limitations}

During our work with production and research applications, we have encountered certain limitations associated with FSDP. This section aims to discuss two tricky caveats that are not readily apparent and pose significant challenges when it comes to troubleshooting. 

\subsubsection{Mathematical Equivalence} \hfill \break

FSDP cannot ensure that it always achieves the same mathematical equivalence as local training, especially with respect to the optimizer computation. This stems from the fact that the optimizer step operates on the sharded parameters, whose data layout is a function of FSDP's \code{FlatParameter} sharding algorithm that does not respect individual parameter boundaries. As a result, any optimizer computation that depends on an original parameter's unsharded value (\emph{e.g.}~vector norm), its tensor structure (\emph{e.g.}~approximate second-order optimizers), or require global states over all parameters will become invalid. Addressing this requires uneven sharding, padding, or extra communication, all of which hurt performance. Co-designing such optimizer computations with sharding is an open research question.

\subsubsection{Shared Parameters} \hfill \break

For shared parameters, FSDP must ensure to not flatten them into multiple \code{FlatParameter}s and to ensure that they are unsharded properly when needed for all usages. If handled incorrectly, PyTorch may raise an error regarding missing tensor storage or size mismatch, which can happen when an FSDP unit attempts to use a shared parameter that has already been resharded by a preceding FSDP unit. The current recommendation is to construct FSDP units such that the shared parameter belongs to the lowest-common-ancestor unit to ensure that the shared parameter is unsharded throughout all usages. This may require some inspection of the model structure to do correctly and may undesirably keep the \code{FlatParameter} unsharded for a large interval, so we are investigating approaches to improve shared parameter handling.
\section{Conclusion}\label{sec:conclusion}

This manuscript elucidates the underlying rationale, design philosophy, and implementation of \code{FullyShardedDataParallel} as of PyTorch 2.0 release. FSDP attains usability and efficiency through a set of advanced techniques, including deferred initialization, flexible sharding strategies, communication overlapping and prefetching, and rate limiting communication collectives. All of these techniques are closely co-designed with other key PyTorch components to ensure the solution is sound and robust. Evaluations show that FSDP can facilitate large language and recommendation models with near linear scalability. 

\begin{acks}
We are grateful to the PyTorch community and PyTorch FSDP users for their feedback and contributions.
\end{acks}

%\clearpage

\bibliographystyle{ACM-Reference-Format}
\balance
\bibliography{references}
\end{document}